\documentclass[journal=jacsat,manuscript=article]{achemso}
\SectionNumbersOn
\usepackage[version=3]{mhchem} 

\usepackage[normalem]{ulem}
\usepackage{comment}
\usepackage[usenames,dvipsnames]{xcolor}

\newcommand{\CX}{\ensuremath{\mathcal{X}}}

\author{Kevin Rossi}
\affiliation{Laboratory of Computational Science and Modeling (COSMO), Institute of materials, Ecole Polytechnique F\'{e}d\'{e}rale de Lausanne (EPFL), Lausanne, 1015, Switzerland.}
\altaffiliation{The authors contributed equally to this work.}
\author{Veronika Jur\'{a}skov\'{a}}
\affiliation{Laboratory for Computational Molecular Design (LCMD), Institute of Chemical Sciences and Engineering, Ecole Polytechnique F\'{e}d\'{e}rale de Lausanne (EPFL), Lausanne, 1015, Switzerland.}
\altaffiliation{The authors contributed equally to this work.}
\author{Raphael Wischert}
\affiliation{~Eco-Efficient Products and Processes Laboratory, Solvay, RIC Shanghai, China}
\author{Laurent Garel}
\affiliation{Aroma Performance Laboratory, Solvay, RIC Lyon, France}
\author{ Cl\'{e}mence Corminb\oe uf}
\affiliation{Laboratory for Computational Molecular Design (LCMD), Institute of Chemical Sciences and Engineering, Ecole Polytechnique F\'{e}d\'{e}rale de Lausanne (EPFL), Lausanne, 1015, Switzerland.}
\altaffiliation{National Centre for Computational Design and Discovery of Novel Materials
(MARVEL), \'Ecole Polytechnique F\'ed\'erale de Lausanne, 1015 Lausanne, Switzerland}
\email{clemence.corminboeuf@epfl.ch}
\author{ Michele Ceriotti}
\affiliation{Laboratory of Computational Science and Modeling (COSMO), Institute of materials, Ecole Polytechnique F\'{e}d\'{e}rale de Lausanne (EPFL), Lausanne, 1015, Switzerland.}
\altaffiliation{National Centre for Computational Design and Discovery of Novel Materials
(MARVEL), \'Ecole Polytechnique F\'ed\'erale de Lausanne, 1015 Lausanne, Switzerland}
\email{michele.ceriotti@epfl.ch}

\title[An \textsf{achemso} demo]
  {Simulating solvation and acidity in complex mixtures with first-principles accuracy: the case of CH$_3$SO$_3$H and H$_2$O$_2$ in phenol}


\begin{document}


\begin{abstract}
We present a generally-applicable computational framework for the efficient and accurate characterization of molecular structural patterns and acid properties in explicit solvent using H$_2$O$_2$ and CH$_3$SO$_3$H in phenol as an example.
In order to address the challenges posed by the complexity of the problem, we resort to a set of data-driven methods and enhanced sampling algorithms. The synergistic application of these techniques makes the first-principle estimation of the chemical properties feasible without renouncing to the use of explicit solvation, involving extensive statistical sampling.
Ensembles of neural network potentials are trained on a set of configurations carefully selected out of preliminary simulations performed at a low-cost density-functional tight-binding (DFTB) level. Energy and forces of these configurations are then recomputed at the hybrid density functional theory (DFT) level and used to train the neural networks.
The stability of the NN model is enhanced by using DFTB energetics as a baseline, but the efficiency of the direct NN (\textit{i.e.}, baseline-free) is exploited via a multiple-time step integrator.
The neural network potentials are combined with enhanced sampling techniques, such as replica exchange and metadynamics, and used to characterize the relevant protonated species and dominant non-covalent interactions in the mixture, also considering nuclear quantum effects.
\end{abstract}


\section{Introduction}

The computational description of structural patterns and acidity constants in condensed molecular environment is a challenging problem.
It relies upon both accurate \textit{ab initio} quantum chemistry (\textit{e.g.}, hybrid density functional theory (DFT)) and converged statistical sampling involving Born-Oppenheimer molecular dynamics.
Yet, the numerical evaluation of energies and forces at the hybrid DFT level is too computationally demanding to achieve long timescale simulations of explicitly solvated species.
\cite{Sulpizi2010, Mangold2011, cheng2014, DeMeyer2016, Gittus2018}
Structural and chemical properties of acids and bases are routinely measured experimentally but their estimate for strong or weak acids as well as unstable species and molecules with multiple tautomeric equilibria is not straightforward.

\begin{figure}[!tbh]
    \centering
    \includegraphics[width=12cm]{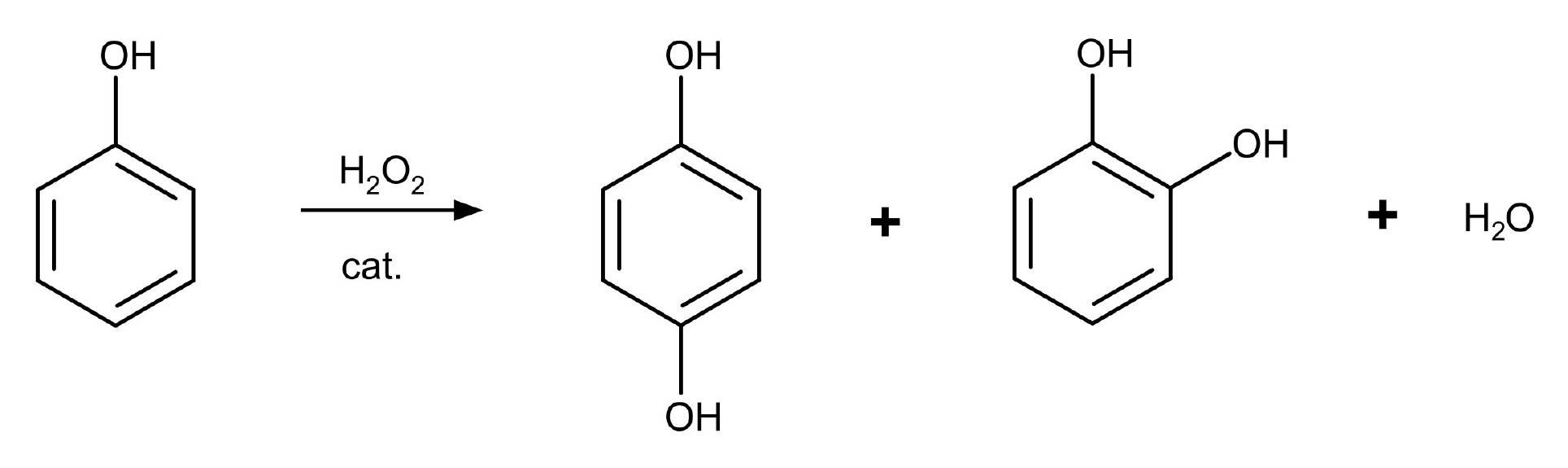}
    \caption{Reaction scheme of the acid-catalyzed hydroxylation of phenol by hydrogen peroxide to form catechol or hydroquinone and water.
    }
    \label{fig:reaction}
\end{figure}

Protonated hydrogen peroxide is a prototypical example that serves as a strong oxidizing agent in acidic media. It is, for instance, used for the catalytic hydroxylation of phenol, as an industrial route towards the production of catechol and hydroquinone, key chemicals in the manufacturing of cosmetic, pharmaceutical, and agrochemical products.\cite{doi:10.1002/14356007.a19_313}
The reaction (see Fig. \ref{fig:reaction}) involves protonated hydrogen peroxide - formed \textit{in situ} in the presence of an acid - that reacts with phenol via electrophilic aromatic substitution.\cite{Varagnat1976}
One of the possible acids for this reaction is methanesulfonic acid (CH$_{3}$SO$_{3}$H, pK$_a$(H$_2$O) = -1.9), which is a non-volatile liquid at ambient temperature, soluble in organic solvents and ambiphilic media.\cite{A900157C}

Identifying the protonation state of methanesulfonic acid and its solvation shells in mixtures of phenol and hydrogen peroxide is complicated by the ambiphilic character of phenol. Phenol acts as a weak acid, which causes a decrease of the common acid strength in comparison to the aqueous medium.
CH$_{3}$SO$_{3}$H interacts with phenol through both hydrogen bonding and apolar interactions owing to presence of the hydroxyl group and an aromatic ring.
From the computational perspective, achieving an accurate description of the non-trivial interplay between hydrogen bonds and/or $\pi$-interactions between the acid and hydrogen peroxide should help rationalizing the stability of the possible reaction intermediates and even the regioselectivity of the reaction.

Here, we demonstrate how the combination of data-driven and enhanced sampling techniques helps characterizing molecular patterns in ambiphilic media. First, we develop a set of neural-network-based reactive force fields, which retain (in the interpolative regime) the accuracy of the hybrid DFT energy and force computations they are trained on, at a fraction of their computational cost.
We train both a baseline neural network (NN) correction that promotes semiempirical DFTB to a DFT hybrid accuracy, and a somewhat less robust albeit much faster ``direct'' neural network that reproduces the potential energy surface of the reference model. 
The trained neural networks (\textit{i.e.,} baselined and direct) are integrated into a molecular dynamic driver, making use of a multiple time-stepping (MTS) approach, to perform replica exchange molecular dynamics (REMD), metadynamics and path integral MD (PIMD). Finally, we use data-driven analysis techniques to characterize and estimate the occurrence of recurrent structural patterns in the solvation environment of the different species. 

The article starts with a general description of the proposed workflow, followed by technical details associated with the theory and implementation of the reference forces and energies, the training of the neural network, and the enhanced sampling schemes. The framework is then validated on a test set, and applied to the analysis of the solvation of hydrogen peroxide and methanesulfonic acid in phenol, taken as an illustrative example. 

\begin{figure*}[t!]
    \centering
    \includegraphics[width=16cm]{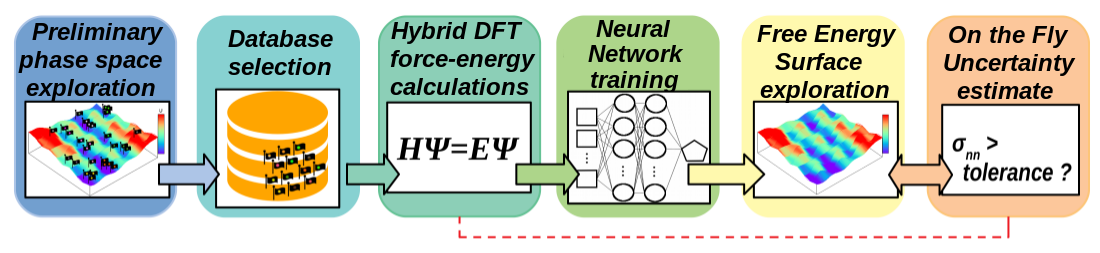}
    \caption{Graphical summary of the essential ingredients of the workflow we used to train a robust MLP, and to use it to sample the free-energy landscape of a complex solution }
    \label{fig:workflow}
\end{figure*}

\section{Operational Workflow}

We begin by providing an overview of the workflow we introduce to achieve an accurate yet computationally affordable exploration of the free energy landscape of a complex, fully-solvated chemical system. The workflow is illustrated in Fig. \ref{fig:workflow}, and is based  on the following steps:

\begin{itemize}

\item {\bf Preliminary phase space exploration:} 
A relevant portion of the phase space is explored at a low-cost electronic structure level (\textit{e.g.}, semi-empirical method).

\item  {\bf Database selection:} A set of distinct structures is selected by a farthest point sampling (FPS) algorithm to avoid structural redundancy. 

\item  {\bf Reference energies and forces:} The forces and energies of the configurations selected at the previous step are computed at the reference electronic structure level (\textit{e.g.}, beyond DFT or hybrid DFT including London dispersion corrections).

\item  {\bf Training of the Neural Networks:}  A NN model is trained to reproduce DFT energies and forces, both directly, and using the semiempirical method as a baseline \cite{Ramakrishnan2015, Sun2019}.

\item  {\bf Free energy surface exploration:} Extensive sampling is performed by combining direct and baselined NNs in MD simulation runs integrated through a MTS scheme \cite{tuck+92jcp} to achieve the accuracy of the latter and exploit the efficiency of the former.

\item  {\bf On-the-fly uncertainty estimate:} On-the-fly uncertainty estimations based on committee models \cite{behler2016, peterson2017, musil2019, janet2019} are used to monitor the extrapolation error and, if necessary, feed novel structures to the NN, to improve its accuracy and reliability along the simulations.
\end{itemize}

\section{Methods}
\subsection{Machine Learning Potentials}
\label{ss:nn}
 
Machine learning potentials (MLP) trained on DFT data are increasingly used to achieve fast-and-accurate prediction of molecular energetics involved in complex atomistic systems.
They have often been employed to investigate chemical reactions in the gas phase \cite{RochaFilho2003, Lorenz2006, Handley2010, Chen2018, Brorsen2019, Amabilino2019, Hase2019}, and the properties of  materials~\cite{Behler2007,bart+10prl,eshe+12prl,soss+12prb},
but recent focus has also been placed on reactions occurring in aqueous media.\cite{ Brickel2019,Hellstrom2018,Hellstrom2016,Hellstrom2017,Shen2018,Schran2019}
Specifically, mechanisms and energetics associated with proton transfer in aqueous system have been investigated for the case of zinc-oxide water interfaces \cite{Hellstrom2018}, Na$^{+}$\cite{Hellstrom2016, Hellstrom2017} or glycine \cite{Shen2018} solvated in water, as well as for water clusters \cite{Schran2019}.
Many frameworks have been proposed to construct machine learning potentials.\cite{Behler2007, Bartok2015, Glielmo2018, Chmiela2017}
These differ by the choice of the atomic structure representation and by the regression scheme. 
\paragraph*{Atomic symmetry functions and Neural Network potentials}

Here we employ Behler-Parrinello atomic symmetry functions (ASF) \cite{Behler2007} as an input to a feed-forward neural network. The global energy $E(\mathcal{R})$ is approximated as the sum of local atomic energy contributions $\epsilon(\mathbf{q}^i_{ASF})$
\begin{equation}
E(\mathcal{R}) = \sum_i \epsilon(\mathbf{q}^i_{ASF})  \mbox{~,}
\label{eq:global_energy1}
\end{equation}
The local energy $\epsilon(\mathbf{q})$ is expressed as a two-layers feed-forward neural network (NN)~\cite{behl11pccp}, whose parameters are optimized to minimize the error on an appropriately constructed training set.  
%
%
The feature vector $\mathbf{q}^i_{ASF}$ is built to provide a symmetry-invariant representation of the environment of the $i$-th atom, and contains atom-centered symmetry functions (ASF) $G_2$ and $G_3$, defined as in Ref.~\citenum{behl11jcp}.
The choice of a reasonably complete, yet non redundant set of ASFs is one of the most delicate and time-consuming aspects in the construction of a Behler-Parrinello  style MLP. Here we automate this selection using CUR decomposition as described in Ref.~\citenum{Imbalzano2018}.
We start from a large set of 192 2-body and 800 3-body symmetry functions per element, with parameters  that span evenly distances up to $\sim$ 7~\AA~ and angles from 0 to 360$^\circ$.
We then use CUR decomposition to select the most descriptive of these functions, choosing 64 for each element. 

\paragraph*{Neural Network simulations}

An MLP trained only on configurations sampled across an \textit{ab initio} trajectory at standard temperature and pressure would often fail when a thermal fluctuation generates highly distorted structures (\textit{e.g.}, atoms in close contact), since it would then enter an extrapolative regime for which no regularizing effect enforces short-distance repulsion. The use of a physics-informed surrogate model as a baseline prevents this possible shortcoming and facilitates the learning of forces and energy predictions. \cite{Ramakrishnan2015, Sun2019}

The baselined NN is however significantly more computationally demanding because of the cost of performing a baseline level computation at each simulation step. An additional $n$-fold gain in speed is achieved by integrating the dynamics via a multiple-time-stepping scheme, in which direct neural network predictions are corrected every $n^{th}$ step by the baselined NN estimate.  

\paragraph*{Uncertainty Quantification}
\label{sec:uncertainty}

Machine learning potentials yield accurate out-of-sample predictions only when used in an interpolative regime. 
To probe whether the neural network predictions take place in an extrapolative regime, it is useful to have a scheme that provides uncertainty estimates.\cite{behler2016, peterson2017, musil2019, janet2019}
Here we use a scheme based on a committee model, \textit{i.e.}, an ensemble of $M$ neural networks that are trained on different subsets of the training set. The additional cost connected with the use of multiple models is offset by the increased reliability afforded by a model with uncertainty quantification. Furthermore, the use of an ensemble based on the same symmetry functions and architecture could allow a substantial reduction of the overhead, which however we do not exploit due to limitations of the current implementation.
The average of the force/energy predictions for a configuration $\CX$, $\bar{y}(\CX)=\sum_i y^{(i)}(\CX )/M$ is taken as the best estimate, and is used to drive the dynamics; the standard deviation across the committee, $\sigma(\CX)$, is taken as a qualitative measure of the uncertainty.
In order to improve the quantitative accuracy of such estimate, the standard deviation is scaled by a factor $\alpha$, ${\sigma(\CX)} \leftarrow \alpha \sigma(\CX)$,
that is determined by maximizing the log-likelihood of the predictive distribution over a validation set of size N$_{val}$:
\begin{equation}
    \alpha^{2} = \frac{1}{N_{val}} \sum_{n}\frac{(y_n-\bar{y}(\CX_n))^{2}}{\sigma(\CX_{n})^{2}}  \mbox{~.}
    \label{eq:uq-alphaa}
\end{equation}
where $y_n$ indicates the reference value for the input $\CX_{n}$.\cite{musil2019}

\subsection{Phase space exploration and characterization}

In addition to the computational gain provided by the machine learning potentials, enhanced sampling MD schemes allow for the efficient exploration of complex free energy landscapes, thus providing insights on species relative stability and kinetics in \textit{e.g.}, proton transfer reactions.\cite{JungMeePark2006, Jung-GooLee2006, Tummanapelli2014, PerezdeAlbaOrtiz2018, Sakti2018, Arunachalam2019, Grifoni2019, Daub2019}
The combination of MLPs and enhanced sampling techniques such as metadynamics\cite{Laio2008} (MetaD) makes the computational exploration of reactions in explicit solvents possible even over extended time and length scales.

The long time-scale trajectories that can be achieved pose an additional challenge when it comes to identify the most relevant species, molecular motifs and reactive events. 
Fortunately, data-driven techniques provide a framework that can also be used for unbiased structural characterization\cite{Ceriotti2019,Engel2018,Helfrecht2019,Vannay2018,Sawatlon2019,Geiger2013, Fulford2019}.
Here we use the sketch-map dimensionality reduction algorithm~\cite{Ceriotti13023,Tribello2012,Ceriotti2013}. Similar to multi-dimensional scaling~\cite{cox-cox10book}, sketch-map tries to find a low-dimensional representation of a set of configurations, matching the distances between high-dimensional sets of features that describe each structure, and those between their projections. A non-linear transformation of the distances helps disregarding uninteresting features (\textit{e.g.}, thermal fluctuations) and obtaining a map in which recurring structural motifs are clearly identified as separate clusters. 

\section{Computational details}

After having summarized the overall methodological framework, we give specific details of the reference electronic structure computations, the training of the MLP and the statistical sampling strategy.

\paragraph*{Quantum chemistry}

DFTB3 with 3OB parameters\cite{Gaus2013,Gaus2014} and the D3H5 \cite{Rezac2017} correction is used as a baseline, which is robust enough to avoid completely unphysical configurations, but is known to occasionally yield qualitatively incorrect predictions, \textit{e.g.} a planar equilibrium structure for gas-phase \ce{H2O2}.\cite{doi:10.1063/1.1871913,Gaus2013}
DFTB computations are performed with the DFTB+ 18.2 software interfaced with the dynamic driver i-PI.\cite{ B.Aradi2007, Gaus2011, Gaus2013, Gaus2014,Rezac2017,Petraglia2016}
The reference energies and forces used for the training are obtained at the PBE0\cite{Adamo1999,Ernzerhof1999}-D3BJ\cite{Grimme2011} level as implemented in CP2K 6.1.\cite{Hutter2014}
All elements are described with the TZV2P-MOLOPT basis set \cite{VandeVondele2007} with cores represented by the dual-space Goedecker-Teter-Hutter pseudopotentials (GTH PBE).\cite{Krack2005}
The plane-waves cut-off is set to 700 Ry with a relative cut-off of 70 Ry. 
All computations employ a Coulomb operator truncated at R = 6 \AA~ and the auxiliary density matrix method with a cpFIT3 fitting basis set. \cite{Guidon2010}
We use converged PBE-D3BJ wave functions as the initial guess for the PBE0-D3BJ computations. We evaluate the energies and forces of 3304 carefully selected configurations (see next section).

\paragraph*{Training set}
The various mixtures used to train the NN comprise 20 phenol molecules, one methanesulfonic acid molecule, up to four water and hydrogen peroxide molecules. 
A set of 3048 configurations is selected from REMD trajectories performed at the DFTB level (see subsection \ref{s:md} for further details) using the FPS scheme relying upon Hausdorff distances in the metric described by the symmetry functions used for the neural network.\cite{Imbalzano2018}
256 additional structures are extracted from high temperature (600~K), high pressure (above ~1000 atm), and PIMD simulations such as to expand the NN training set with a small albeit informative number of highly distorted configurations.

\paragraph*{Neural network training}

The NNs are trained with 2 layers and 22 nodes per layer to predict PBE0-D3BJ level forces and energies. The weights in the NN are optimized via Kalman Filtering routines.
We use 100\% of the energies and 0.5\% of the force components per configuration in the training fraction, with force weights 8-fold larger relative to energy weights. The ASF calculation and NN training is performed with the n2p2 package.\cite{Singraber2019a}
Over-fitting is avoided by using an early stopping criterion. A total of 400 training iterations is allowed.
An ensemble of 5 neural networks is used to compute the baselined force and energy predictions. 
The uncertainty calibration for the energy predictions carried out by the baselined NN yields $\alpha=5.8$.

\paragraph*{Molecular Dynamics}
\label{s:md}

Preliminary replica exchange molecular dynamics (REMD) simulations are performed with 8 constant-temperature replicas (333 K, 339 K, 347 K, 358 K, 370 K, 384 K, 403 K, 423 K), initiated from the same structure, with randomly sampled momenta from the appropriate distribution at each temperature. Swaps among replicas are attempted every 10 steps. Simulations were run for 15 - 40 ps per replica (in total 2 368 ps).

Two specific mixtures are considered for further study, using the trained NNs to achieve more thorough sampling. One of the mixtures contains 20 phenol molecules and one methanesulfonic acid molecule, the second also contains one molecule of hydrogen peroxide.

MD trajectories are integrated using i-PI \cite{kapil2019} with the DFTB+ and LAMMPS drivers for forces and energy computations.
Equations of motion are integrated using a multiple time step (MTS) scheme \cite{Tuckerman1992}, with an outer time step of 3~fs (involving a DFTB calculation and a NN correction) and an inner one of 0.5~fs (involving a direct MLP fitted to DFT calculations). We improve the stability of the trajectories using a BAOAB\cite{Leimkuhler2016} splitting, and apply a velocity rescaling thermostat with a frequency of 10~fs, together with a thermostat based on Generalized Langevin Equation (GLE)\cite{ceri+10jctc}. In order to stabilize MTS trajectories without slowing down diffusion, the GLE is designed so that modes above 100 THz are affected by a friction of 125 ps$^{-1}$, while the effective frictions diminishes to 0.01 ps$^{-1}$ for frequencies approaching zero. \cite{Morrone2011}
MD sampling is performed for 720 ps using REMD, with the same temperature distribution used for the preliminary DFTB simulations.
Nuclear quantum effects (NQEs) are investigated for systems at 363~K using data gathered from independent path integral molecular dynamics trajectories for a total of 720~ps. The PIGLET technique~\cite{ceri-mano12prl} is used to reduce the number of replicas needed for convergence. Consistently with previous simulations of molecular liquids at room temperature, 6 path integral beads were found to be sufficient. Generalized Langevin equation parameters were obtained from the GLE4MD website~\cite{gle4md}.

\paragraph*{Collective variables and metadynamics}

We focus on the solvation structure around the acid, on the protonation states of the different species, and on the possibility for proton transfer events.
For this reason, we use two collective variables (CVs) for the accelerated sampling simulations. 
(1) The coordination number (CN) of the sulfonyl group. The CN, which has been frequently applied to promote the sampling of proton transfer reactions \cite{JungMeePark2006, Jung-GooLee2006, Tummanapelli2014, PerezdeAlbaOrtiz2018, Sakti2018, Arunachalam2019, Grifoni2019,Daub2019}, is defined here in terms of a smooth switching function of the distances between the oxygen atoms of the acid and hydrogen atoms of the hydroxyl groups (either of phenol or hydrogen peroxide); the CN is normalized so as to avoid attributing the same H atom to multiple species. (2) We employ also a collective variable corresponding to the minimum distance, D, between the hydroxyl hydrogen and the oxygen atoms in the acid, to characterize the H-bonds between the anion and nearby protons. Detailed definitions of the CVs are discussed in section 2.3 in the SI.

MetaD trajectories (with and without incorporation of NQEs) are carried on by depositing gaussians of height 0.8 kJ/mol and width 0.1 and 0.2 in the CV1 and CV2 space respectively every 36~fs. Well-tempering is enforced by using a Gaussian height damping factor $ \Delta T $ such that the ratio $\frac{ \Delta T + T}{ T}$ is equal to 3.
The parameters of the normalized coordination numbers of the oxygen of the methanesulfonyl hydroxyl group, are set to $p_{0} = 0.9$, $q_{0} = 0.4$, $n=6$, and $m=12$, while the minimum distance between the acid oxygen and the hydrogens bonded to any oxygen is computed with $\delta = 24$ (see the SI for a definition of the order parameters). The metadynamics simulations (100 ps for each mixture simulated classically and including NQEs) are performed with PLUMED version 2.5.1 \cite{TRIBELLO2014604} interfaced with i-PI. A representative plumed input is stored in the plumed nest repository \cite{bono+19nm} at https://www.plumed-nest.org/eggs/20/008/.

\begin{figure*}[tbh]
    \centering
    \includegraphics[width=0.49\textwidth,height=3.cm]{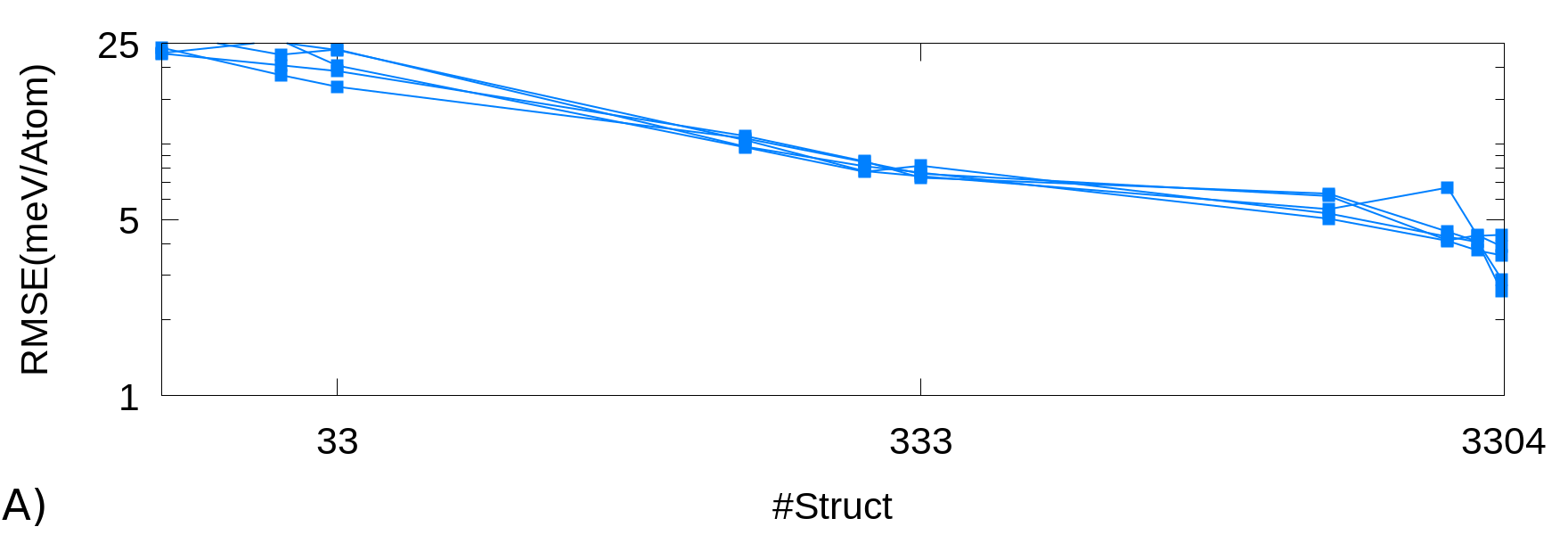}
    \includegraphics[width=0.49\textwidth,height=3.cm]{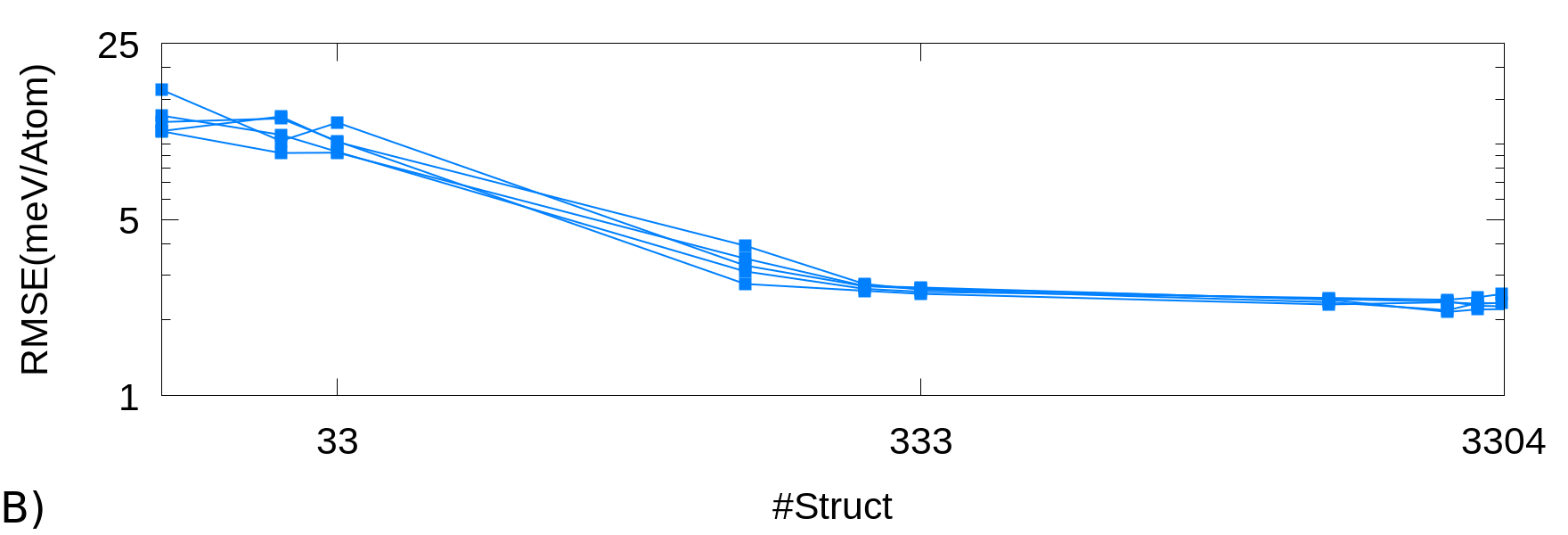}
    \includegraphics[width=0.49\textwidth,height=3.cm]{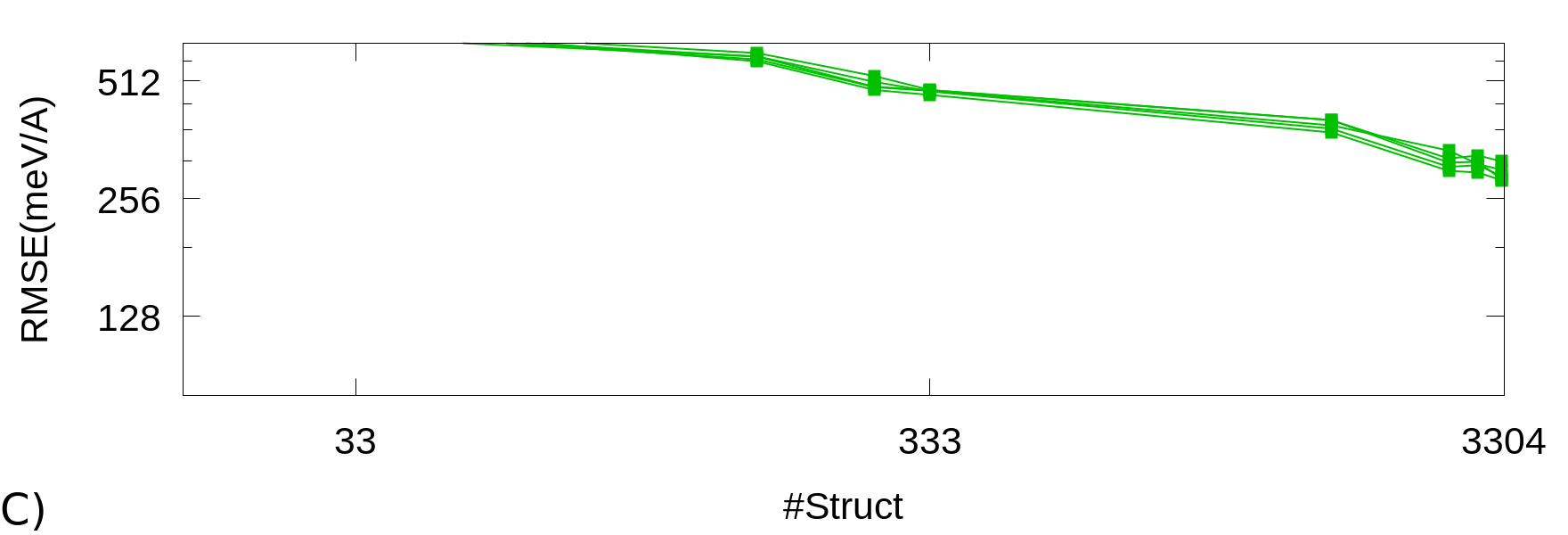}
    \includegraphics[width=0.49\textwidth,height=3.cm]{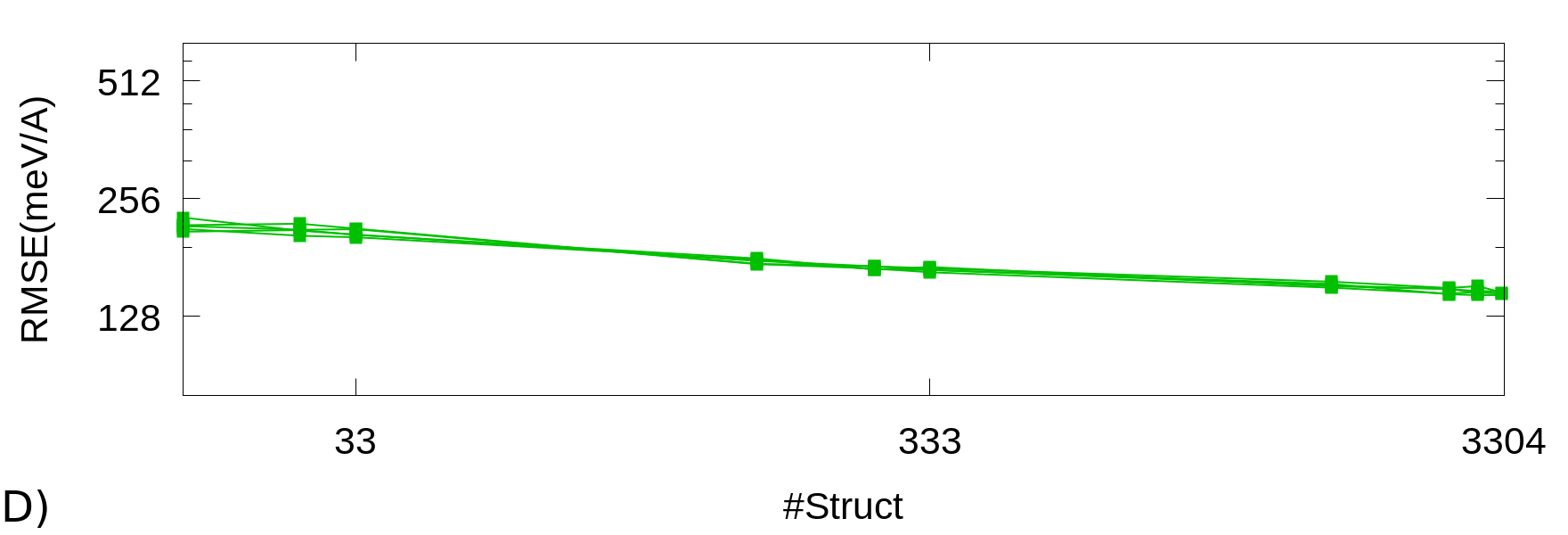}
    \caption{ Energy (direct NN in panel (A), baselined NN in panel (B)) and forces (direct NN in panel (C), baselined NN in panel (D)) learning curves (RMSE vs. relative number of training structures, with the full database comprising 3304 structures). All data are taken after 400 iterations of the neural network loss function optimization, except when the training set comprises 99\% of the structures, for which results after 600 iterations are reported.
    }
    \label{fig:nn-learning-curves}
\end{figure*}

\begin{figure}[tbh]
    \centering
    \includegraphics[width=6cm,height=6cm]{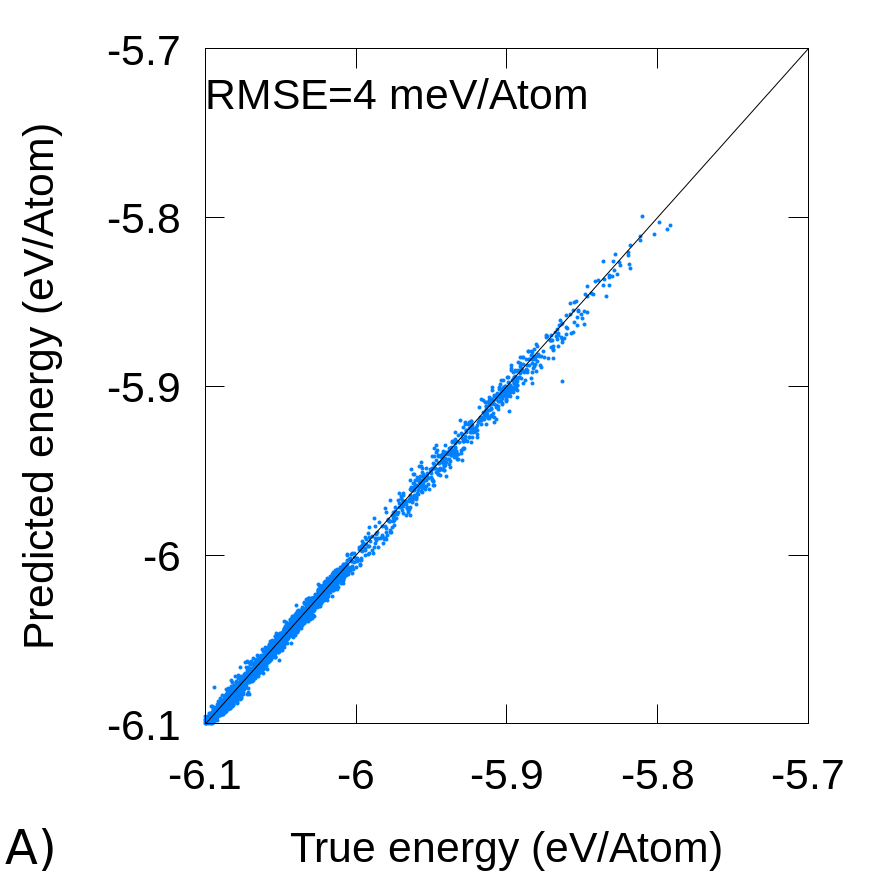}
    \includegraphics[width=6cm,height=6cm]{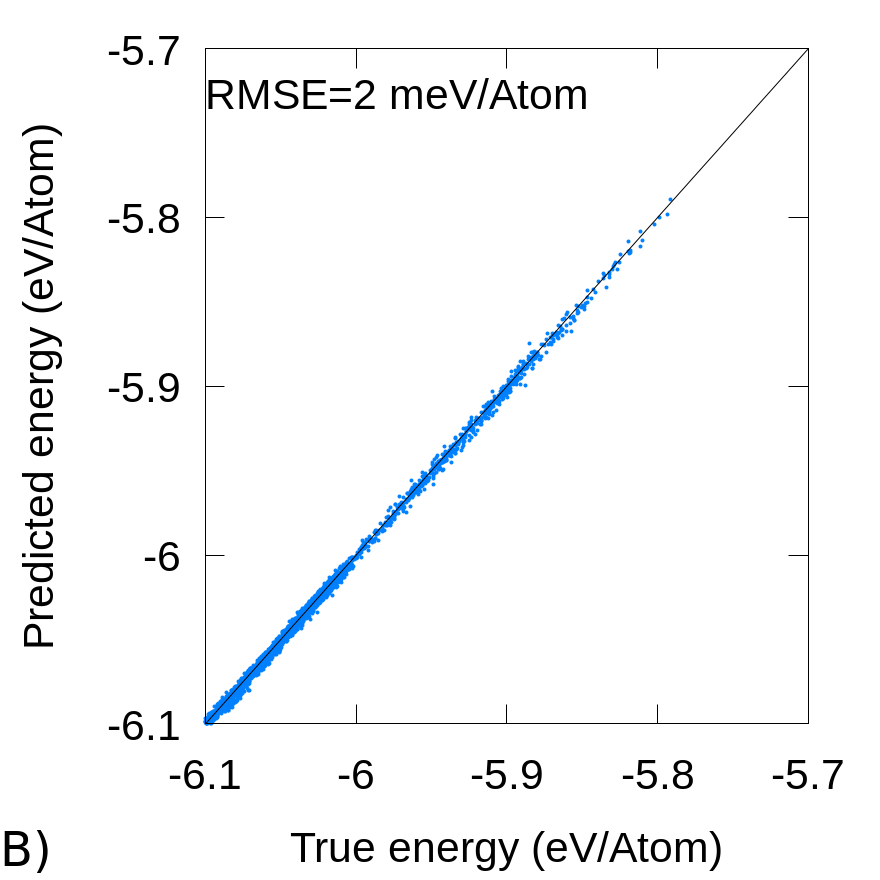}
    \includegraphics[width=6cm,height=6cm]{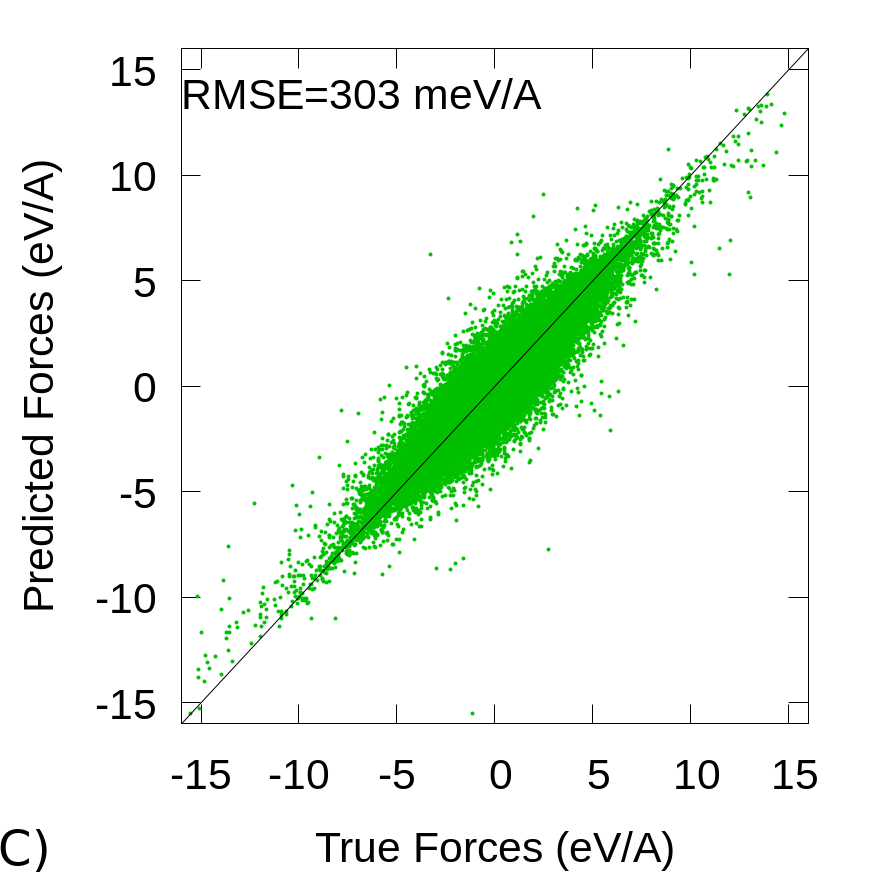}
    \includegraphics[width=6cm,height=6cm]{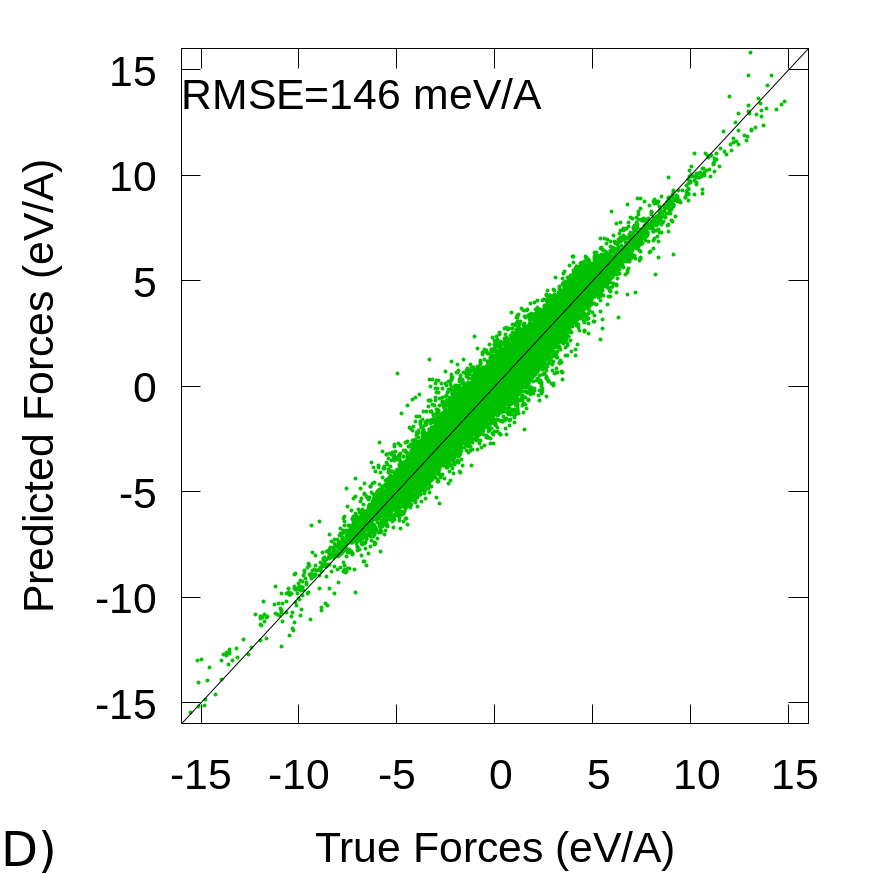}
    \caption{ 
    Parity plots (True vs. Predicted values) for DFTB+baselined NN and direct NN for energies (direct NN in panel (A), baselined NN in panel (B)) and forces (direct NN in panel (C), baselined NN in panel (D)) predictions. Results are reported by taking the average energy and force predictions of the ensembles of  baselined and direct models trained with 80\% of the database structures.
    }
    \label{fig:nn-parity}
\end{figure}

\section{Results}

\subsection{Framework Validation}
\label{ss:nnval}

Before discussing the chemical outcome of the simulations, we present some diagnostics that assess the training performance of the MLP. 
Fig. \ref{fig:nn-learning-curves} reports learning curves for the model, \textit{i.e.}, the root mean square error (RMSE) for the energy (top) and force (bottom) as a function of the number of training structures fed to the neural networks.
The RMSE is computed for a test set of structures randomly extracted from the database described in Section 3.1. 
The left-most panels show curves for the direct MLP, that reach an accuracy of approximately 4 meV/atom for the energy, and 300 meV/\AA{} for the forces, with the largest training set size considered. The power-law decay of the learning curves indicates that the accuracy is data-limited, and that a more accurate model could be obtained, at the price, however, of performing a much larger number of reference computations. 
The baselined model (right panels), on the other hand, achieves easily substantially lower errors (RMSE $<$~3 meV/atom, and $<$~150 meV/\AA{} for energy and forces) and saturates at about 500 training structures, indicating that further increasing the accuracy would require developing a more sophisticated model (\textit{e.g.}, one including long-range interactions) rather than increasing the size of the training set.

\begin{table}[tbh]
    \centering
    \begin{tabular}{cccccc}
         PBE0& PBE & DFTB-D3H5 & DFTB & MTS NN & NN \\
         -D3BJ & -D3BJ  & + NN comm. & -D3H5 & comm. & single \\
         \hline
         \hline
         67$\times 10^3$ & 21$\times 10^3$ & 16 & 12 & 6.7 & 0.8 \\
    \end{tabular}
\caption{CPU time (core seconds) required to advance MD simulations by 0.5 fs. DFT timings (PBE0-D3BJ, PBE-D3BJ) are determined based on computations performed on a single computing node with two 14 cores Intel Broadwell processors running at 2.6 GHz.
DFTB-D3H5 and neural-network force-fields timings are computed based on single-core execution on the same hardware. The ``NN comm.'' label indicates a prediction obtained with a committee of five members, as opposed to ``NN single'' that indicates results from a single member. The MTS cost is computed based on a $\sim$ 40~CPU s timing for a  3~fs outer time step, consisting of 1 DFTB + 35 NN computations (5 DFTB corrections + 30 ``direct'' NNs).  }
    \label{tab:timings}
\end{table}

Parity plots for the direct and baselined-learning with 80$\%$ training set,
are shown in Fig. \ref{fig:nn-parity}.
The direct learning has a RMSE of $\sim$ 4 meV/atom and $\sim$ 300 meV/\AA ~notwithstanding the complexity of the investigated mixture. The baselined model achieves an energy RMSE around 2.5~meV/atom, and a force RMSE below 150~meV/\AA. The lower RMSE is also associated with faster-decaying tails in the error distribution (see Fig. S1 in the SI), underscoring the better stability in comparison with a direct MLP. 

DFTB is more than 1000 times faster than hybrid DFT, for a system of this size, but still about 20 times slower than the evaluation of a single direct NN MLP.
The respective timings of PBE0-D3BJ, PBE-D3BJ, DFTB-D3H5, NN baselined on DFTB-D3H5, and NN computations are gathered in Table \ref{tab:timings}. In order to reduce the cost of a baselined model, we use MTS integration, computing the direct MLP with an inner time step of 0.5~fs, and the DFTB correction at the outer loop. 
Thanks also to the use of a targeted GLE thermostat, that dampens high frequency modes associated with the OH vibrations, the integration is stable up to an outer step of 4~fs (see conserved quantity and temperature time evolution in Fig. S3 in SI). We choose a more conservative value of 3~fs for our production runs. 
The stability of the dynamics is verified in different mixtures comprising 20 phenol molecules, 20 phenol molecules and 1 methanesulfonic acid, and 20 phenol molecules and 1 hydrogen peroxide.
The RDFs obtained from using the MTS scheme were also benchmarked against trajectories obtained by integrating the equations of motion only with the baselined-NN force-field. 
As reported in Fig. S4 in the SI, the RDFs using both approaches are nearly identical. 
The length of the outer step could be further increased by improving the relative accuracy of the direct NN with respect to the baseline corrected potential. 
This comes at the computational cost of refining the direct NN training. 
In addition, such improvement may not be possible $\it{ad ~libitum}$ if the direct NN learning saturates with the number of iterations of the NN optimization or the number of training structures.
Overall, the use of a NN MTS force field leads to a 10$^{4}$ gain in speed with respect to PBE0-D3BJ. 

\subsection{Uncertainty quantification}

\begin{figure}[b!th]
    \centering
    \includegraphics[width=12cm]{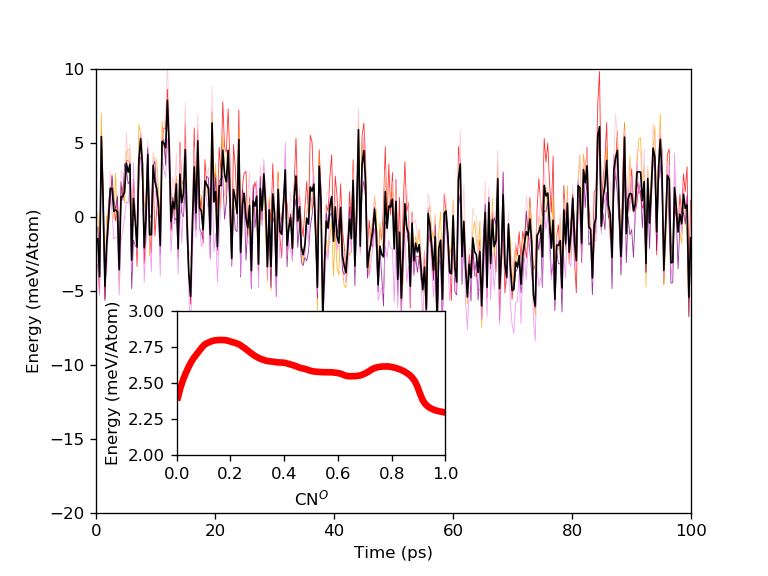}
    \caption{Time evolution of the potential energy during a metadynamics simulation sampling the protonation states of \ce{CH3SO3H}. For each member in the ensemble of neural networks we report the corresponding energy prediction in colored lines. Their average, which is used to propagate the dynamics, is shown in black. The inset shows the mean uncertainty, conditionally averaged by binning over different values of the collective variable CN$^{O}$.}
    \label{fig:uncertainty}
\end{figure}

The use of an ensemble of neural networks for force and energies allows to estimate an uncertainty on the prediction.
Fig. \ref{fig:uncertainty} shows the evolution of the 5 energy predictions for the members of the MLP ensemble (rescaled according to $\alpha$ as discussed in Ref.~\citenum{musil2019}), during one of the classical MetaD trajectories discussed in Section~\ref{sec:uncertainty}.
The spread in the predictions is consistent with a mean estimated uncertainty around 2.5~meV/atom, which is comparable to the accuracy estimated on the validation set. The largest uncertainty observed along the trajectory is of the order of 7 meV/atom, corresponding to a momentary fluctuation rather than to a systematically larger error along an extended section of the trajectory. The inset in the figure shows a conditional average of the uncertainty as a function of the coordination number of the sulfonyl oxygen atoms, that is used to sample the deprotonation reaction. On average, the uncertainty is larger for configurations along the proton dissociation pathway, which correspond to higher free energy values and are poorly represented in the training set. While the predicted error is still very low, $\approx 3$ meV/atom, its magnitude serves to identify under-sampled regions, and could be used in an active learning setting. Thanks to the rather extensive preliminary REMD sampling, however, we did not encounter large-error regions of configuration space that required retraining.

\subsection{CH$_3$SO$_3$H and H$_2$O$_2$ in phenol}

The data from the REMD trajectories (720 ps for each replica) are analyzed addressing three specific questions: 
\begin{itemize}
    \item What is the probability for the acid to be deprotonated? Are proton transfer reactions likely to occur?
    
    \item What are the characteristic features in the H-bond patterns involving the acid, phenol, and hydrogen peroxide?
    
    \item What are the structural signatures associated with apolar interactions, \textit{i.e.}, CH..$\pi$ ?
    
\end{itemize}

\paragraph*{Oxygen environments in a complex mixture}

\begin{figure*}[t!bh]
    \centering
    \includegraphics[width=0.49\textwidth]{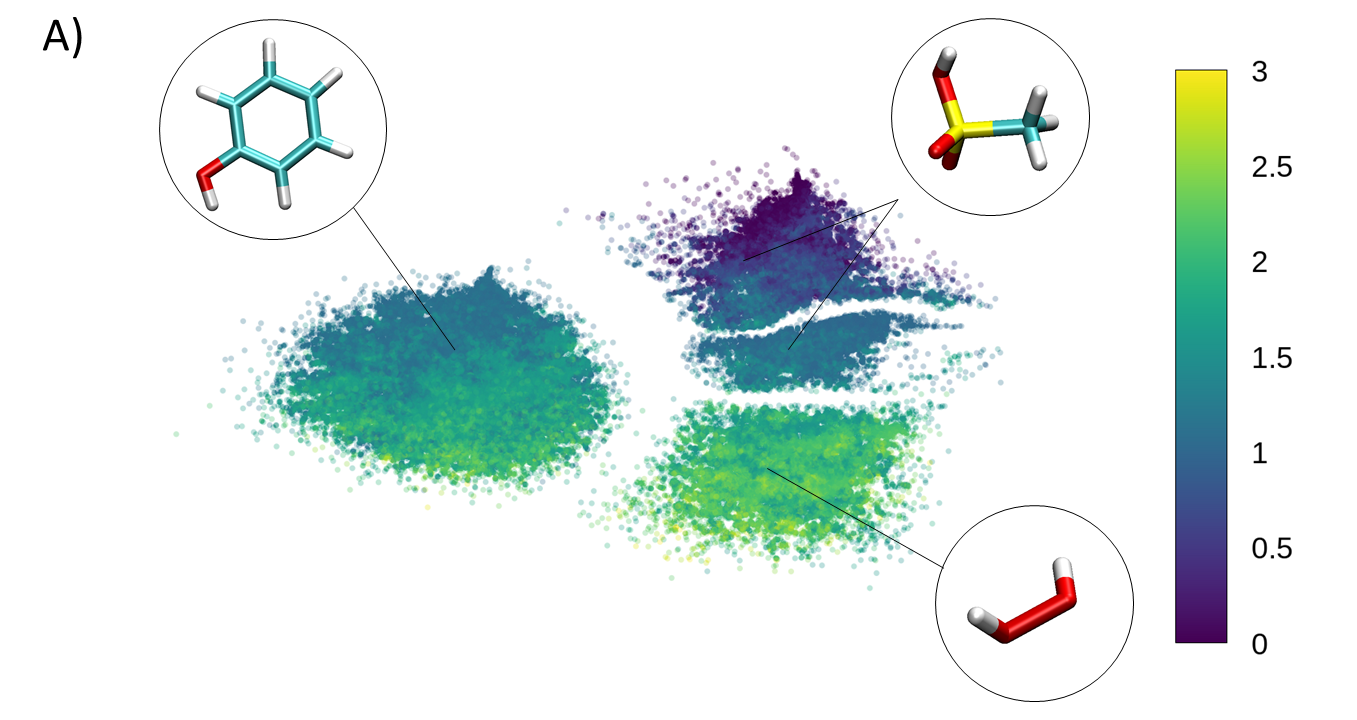}
    \includegraphics[width=0.49\textwidth]{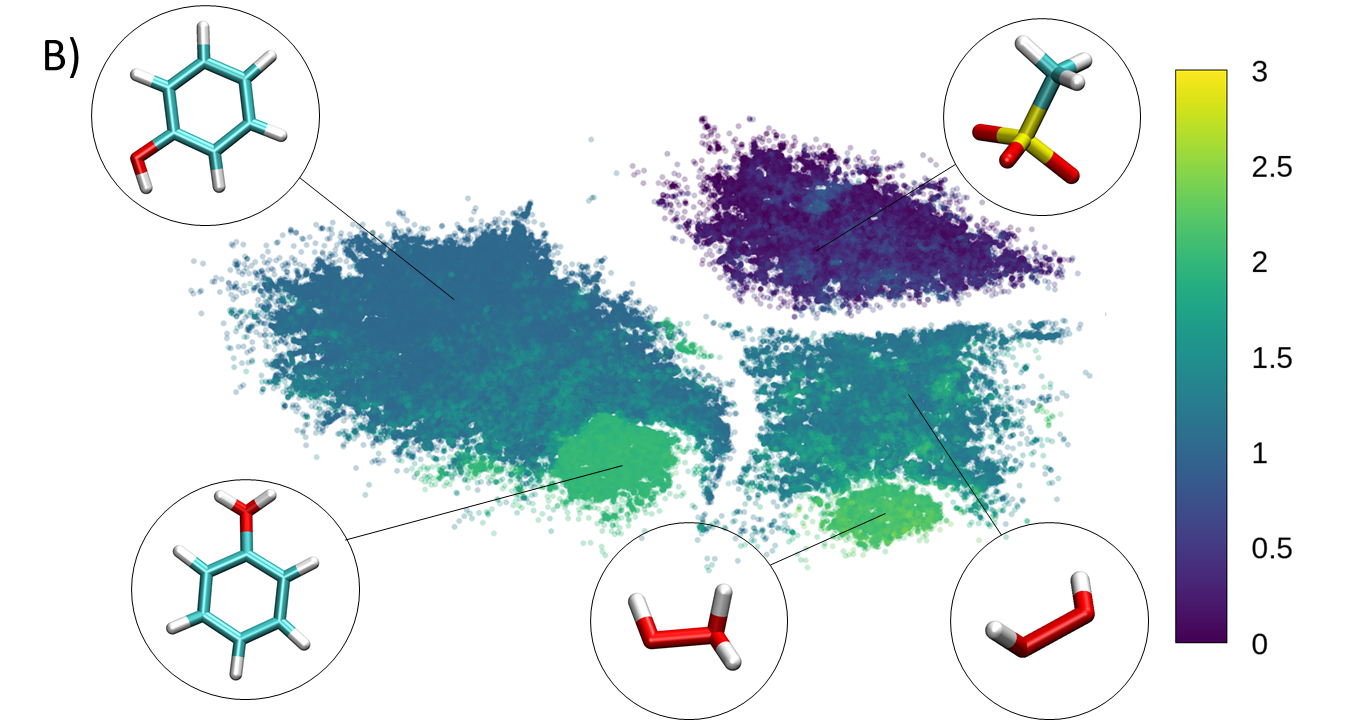}
    \caption{Sketch-map displaying species in the mixture containing 1 H$_{2}$O$_{2}$ for classical REMD sampling carried out using  NN-corrected (A) and DFTB-D3H5 (B) energy and force predictions. 358-K replica sampled for 52 ps (DFTB-D3H5) and 240 ps (NN) are used. Points represent different oxygen atoms. The color code corresponds to the sum of the reciprocal distances of hydrogen atoms within 1.5 \AA~(DFTB) and 2 \AA~(NN) radius. Details on the description of the atomic environments are provided in the SI (Section 2.3). 
    }
    \label{fig:sketchmap_fes}
\end{figure*}

The different species in the mixture are identified and analyzed with a sketch-map representation (Fig.~\ref{fig:sketchmap_fes}) that relies on a set of features describing the atomic environment of the oxygen atoms, making the distinction between the different moieties and their possible protonation states (for further details see Section 2.3 in the SI). 
Fig.~\ref{fig:sketchmap_fes} represents the chemical environment of oxygen atoms in a phenol solution containing one methanesulfonic acid and one hydrogen peroxide molecule at the DFTB-D3H5 and NN-corrected levels.
With the DFT-quality MLP, the methanesulfonic acid does not promote the protonation of any other species and remains protonated along the REMD simulation. This is reflected in the  sketch-map representation by the presence of four clusters shown in Fig. \ref{fig:sketchmap_fes}A. One cluster regroups the phenol oxygen atoms, one is formed by the H$_2$O$_2$ oxygens, whereas the last two corresponds to the OH and sulfonyl group of the acid.
Less accurate DFTB-D3H5 results, on the contrary, favor stable anionic structures, as apparent in the corresponding sketch-map analysis (Fig.~\ref{fig:sketchmap_fes}B) that shows two additional clusters corresponding to \ce{PhOH2+} and \ce{H3O2+}.
Such a discrepancy highlights the necessity of using higher level (\textit{e.g.}, PBE0-D3BJ) energies and forces for capturing the correct qualitative protonation state of the acid.

\paragraph*{Deprotonation free energy}

\begin{figure}[t!bh]
    \centering
    \includegraphics[width=12cm]{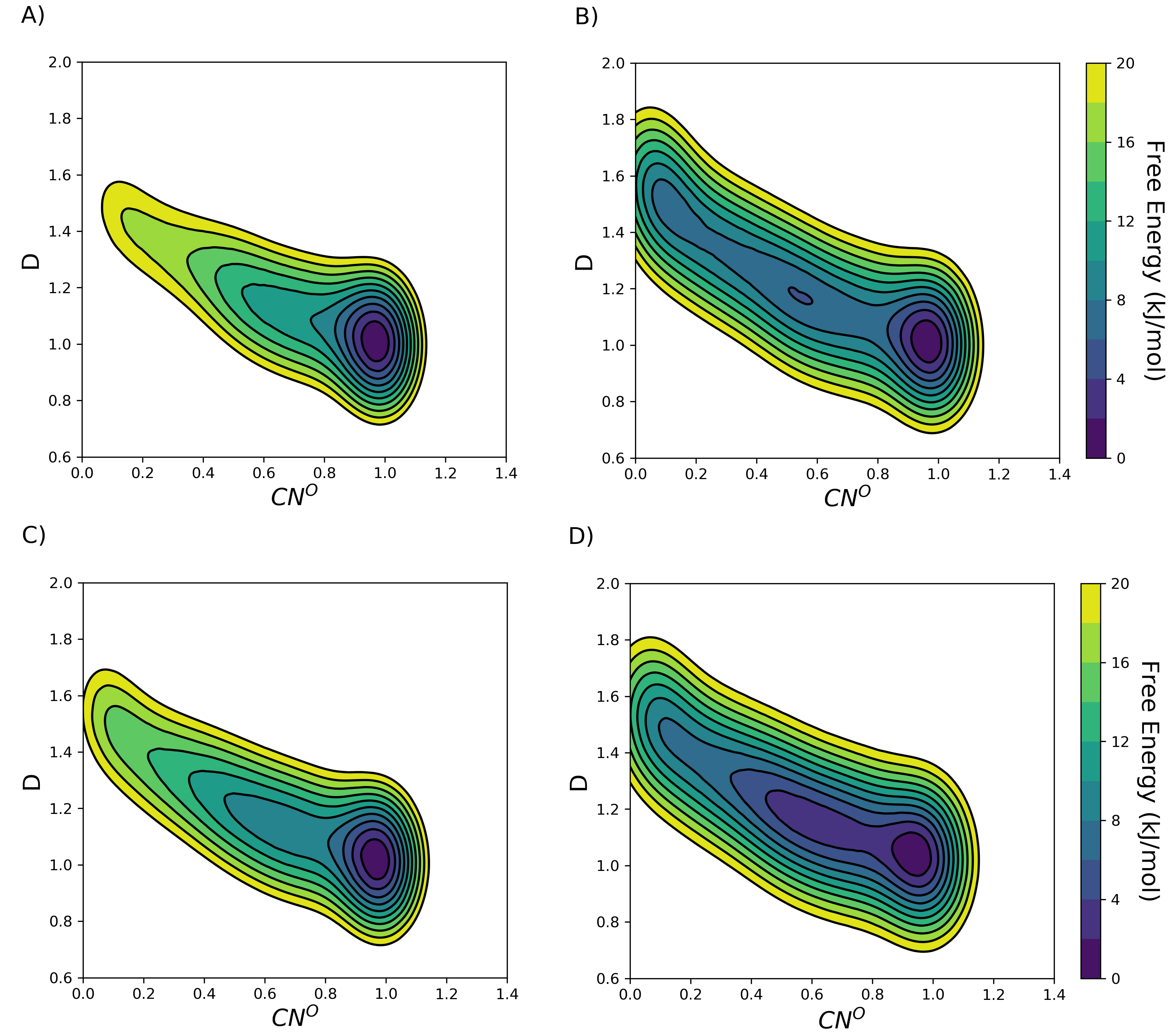}
    \caption{2D Free energy landscape projections without (upper panels) and with NQEs (lower panels) for 1 CH$_3$SO$_3$H in phenol (A, C) and  1 CH$_3$SO$_3$H and 1 H$_{2}$O$_{2}$ in phenol (B, D). The minimum at (1,1) corresponds to CH$_3$SO$_3$H, the deprotonated acid is mapped around CN$^{O}$ $\sim 0$.}
    \label{fig:protonation_fes}
\end{figure}

A quantitative estimate of the pKa of methanesulfonic acid with respect to the protonated phenol and hydrogen peroxide can be extracted from the MetaD sampling at 363~K. Fig. \ref{fig:protonation_fes} A and B show the reconstructed free energy profile for the deprotonation of CH$_3$SO$_3$H in MetaD simulations.
The free energy minimum in the presence of H$_{2}$O$_{2}$ corresponds to the protonated acid, which is about 8 kJ/mol lower than the other species.
In pure phenol, the free energy difference between the neutral and deprotonated acid is almost twice this value.
This comparison highlights that H$_{2}$O$_{2}$ facilitates the deprotonation of CH$_{3}$SO$_{3}$H solvated in phenol, \textit{i.e.}, that the relative acidity of the protonated species is \ce{PhOH2+}$>$\ce{H3O2+}$>$\ce{CH3SO3H}.
Accounting for NQEs by performing path integral MetaD simulations (Fig. \ref{fig:sketchmap_fes} panels C and D) stabilizes the deprotonated species by up to 2 kJ/mol in both mixtures, which corresponds roughly to a change of 0.5 pH units.  
However, this does not change the qualitative picture: the protonated acid remains the most stable species, followed by \ce{H3O2+} and by \ce{PhOH2+}. NQEs would be important to determine quantitatively the pKa values, but they are not as crucial as the use of a high-quality \emph{ab initio} MLP. 

\paragraph*{Characterization of the H-bond network}

All the species present in the mixture are potential hydrogen bond donors or acceptors but some bonding patterns are more frequent than others. 
The focus is especially placed on the H-bond network involving the acid. In its protonated form, the acid carries two types of oxygen atoms - the sulfonyl oxygens, which are hydrogen bond acceptors, and the hydroxyl group which can both donate and accept H-bonds. 

\begin{figure}[!t]
\centering
\includegraphics[width=1.0\textwidth]{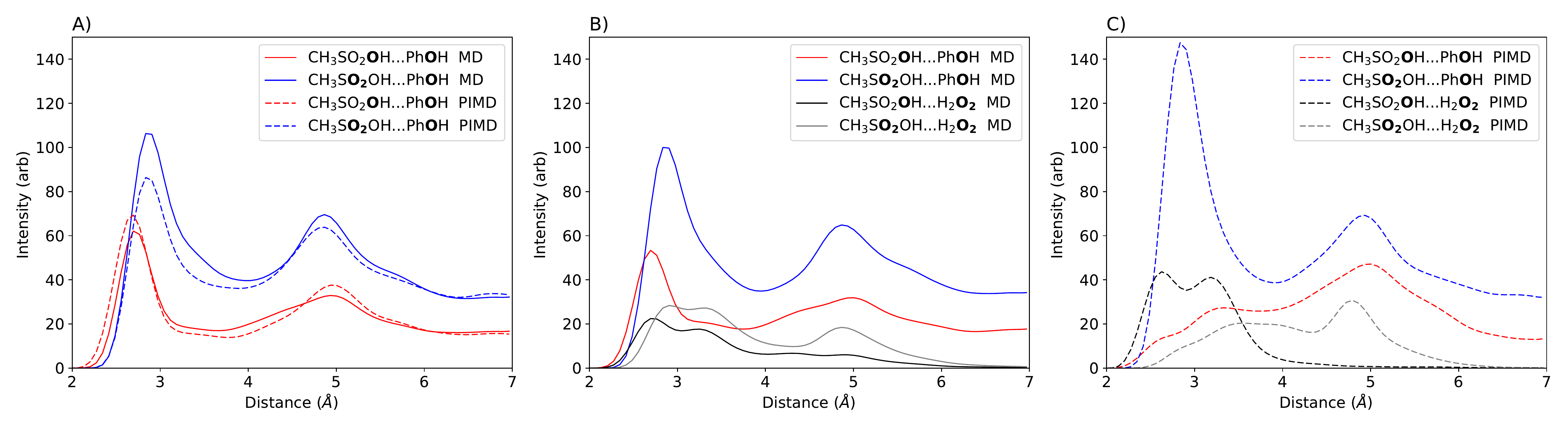}
\caption{
Pair correlation functions between the O atoms of different species in the mixture, indicated by bold face in the legend. The various $g(r)$ are normalized in such a way that they can be meaningfully compared and combined: for instance the sum of the $g(r)$ between the acid hydroxyl and phenol and between the acid hydroxyl and hydrogen peroxide corresponds to the $g(r)$ between the acid \ce{OH} and any other O in solution.  Panel (a) shows the acid-phenol distributions for a simulation box containing only phenol and the acid, and compares classical (358K) and PIMD (363K) trajectories. Panel (b) shows $g(r)$ from classical MD of a box containing the acid, phenol, and one \ce{H2O2} molecule. Panel (c) show the pair correlations from the same mixture, sampled quantum mechanically. Based on the analysis of the 3D distribution in Fig.~\ref{fig:SDF}, the first peak in the distributions involving \ce{CH3SO2}{\bf O}\ce{H} and either phenol or hydrogen peroxide can be interpreted as arising from the acid donating an HB, while the distributions involving sulfonyl oxygens are associated to HBs accepted by the acid. 
}
\label{fig:rdf-oh}
\end{figure}

\begin{figure*}[t!b]
    \centering
    \includegraphics[width=15.5cm]{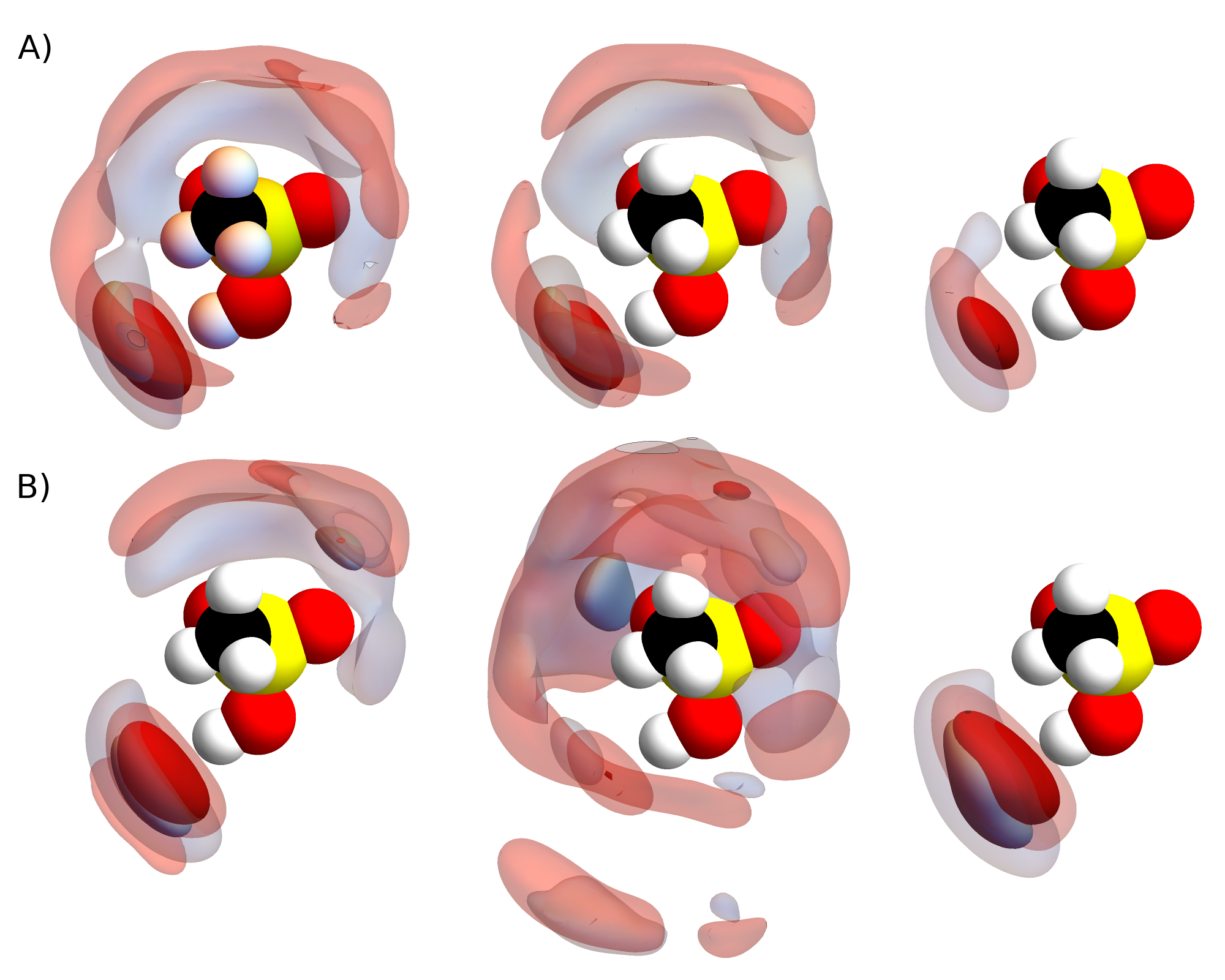}
    \caption{Density distribution of oxygen (red contour) and hydrogen (grey contour) atoms around the acid: Ph{\bf{OH}} in pure phenol (left) and in a mixture of phenol and one H$_{2}$O$_{2}$ (middle); {\bf H$_{2}$O$_{2}$} in the same mixture of phenol and hydrogen peroxide (right).
    Panel A corresponds to the classical REMD simulation, while panel B corresponds to path integral MD, and incorporate nuclear quantum effects. The density isocontours correspond to density values of 0.004 (transparent) and 0.012 (opaque). The density histogram is accumulated in a reference frame defined by the \ce{SOH} atoms: given the flexibility of methanesulfonic acid, the sulfonyl oxygens are not fixed in the ideal positions (see also Fig. S4), contributing to the larger spread of the density in the direction of the \ce{SO2} group compared to that in the direction of the acid hydroxyl, which defines the orientation of the axes. Distributions computed in a reference frame defined by the sulfonyl oxygens are reported in the SI, Fig. S5 and S6. The interested reader is referred to the Supplementary Information Section 5 for a detailed discussion on the degree of localization of the atoms in the acid and the density distribution calculation.}
    \label{fig:SDF}
\end{figure*}

Fig.~\ref{fig:rdf-oh} reports the pair correlation functions $g(r)$ between the O atoms of different species.
Methanesulfonic acid acts both as a HB acceptor and as a donor, with the sulfonyl oxygens behaving as acceptors, and the methanesulfonic hydroxyl group behaving primarily as a donor, as indicated by the 3D distribution in Fig.~\ref{fig:SDF}. Thus, even if $g(r)$ only reports on the correlations between oxygens, without explicit information on hydrogen positions, the peaks around 3~\AA~ can be interpreted in terms of the strength and populations of hydrogen bonds.  
Panel (a) in Fig.~\ref{fig:rdf-oh} shows that NQEs strengthen the HB donated by the acid \ce{OH}, increasing slightly the height of the peak and shifting it towards smaller distances, and weaken the HB accepted by the sulfonyl oxygens. It has been observed consistently that NQEs tend to strengthen strong HBs and weaken weak HBs~\cite{li+11pnas}, suggesting that the HBs donated by the acid are stronger than those it accepts. Panel (b) shows that in classical REMD simulations \ce{H2O2} binds strongly to the acid hydroxyl group, remaining within $\approx$~6\AA{} for the whole trajectory. \ce{H2O2} competes with phenol for binding, reducing the height of the corresponding peak by $\approx$ 15\%. The \ce{CH3SO2OH}--\ce{H2O2} RDFs show a split first-neighbor peak, demonstrating that the acid binds preferentially to one of the two O atoms in the hydrogen peroxide molecule, with the second O staying further apart.
\ce{H2O2} can also donate HBs to the sulfonyl groups, but remains at a somewhat larger distance than phenol, and the $g(r)$ between \ce{PhOH} and the sulfonyl oxygens remain largely unchanged.
Panel (c), that shows the same mixture as in panel (b), modeled by PIMD simulations, demonstrates that NQEs have a substantial impact. The binding of \ce{H2O2} with the acid hydroxyl group is greatly enhanced, with the phenol acceptor peak almost completely suppressed. At the same time, one observes a substantial increase of the peak that corresponds to phenol donating a HB to the sulfonyl oxygens, indicating that the combined effect of the presence of \ce{H2O2} and of nuclear quantum effects changes the structure of the solvation shell around \ce{CH3SO3H}. 

The hydrogen bond patterns discussed in the previous paragraphs are visualized more explicitly by the 3D density distributions reported in Fig. \ref{fig:SDF}.
The acid hydroxyl group is a strong donor towards both phenol and H$_{2}$O$_{2}$ with a stronger affinity towards the latter. Specifically, the density distribution of H$_{2}$O$_{2}$ oxygen atoms around the hydroxyl group is high and directional (right Fig. \ref{fig:SDF}). A similar localized region is observed for the phenol distribution around the hydroxyl group both in pure phenol and in the phenol/H$_{2}$O$_{2}$ mixture (left and middle Fig. \ref{fig:SDF}). 
Hydrogen peroxide binds almost exclusively by accepting a HB from the acid OH. Although binding occurs preferentially to one of the two oxygens, the distribution is broad, and one also observes bifurcated HBs, in which the acid hydroxyl group is shared between the two oxygen atoms of hydrogen peroxide (see illustrative snapshots in Fig. S7).
The trends observed in the O-O $g(r)$ sampled at 358K are qualitatively consistent with those observed over the 333K-423K temperature range (see Figure S8-S13), which indicates that temperature does not affect substantially the solvation environment of \ce{CH3SO3H}.
Nuclear quantum effects have a relatively minor effect on the distribution in pure phenol, but lead to larger changes in the presence of a peroxide molecule. When including NQEs, the latter binds very strongly to the methanesulfonic acid OH group, displacing completely phenol and triggering a rearrangement of the solvation shell, in which phenol binds more strongly to the sulfonyl oxygen atoms. 
It should be stressed that quantitative convergence of the 3D distribution function is not trivial even with several hundred ps of simulations. The distributions in Fig.~\ref{fig:SDF} are obtained from several independent runs, all showing qualitatively consistent trends. 

\paragraph*{CH-$\pi$ interactions}
Apart from the rich hydrogen bond network, the methanesulfonic acid may form apolar interactions with the aromatic phenol ring through its methyl group. The relevance of CH-$\pi$ interactions between the acid methyl group and phenol ring can be further analyzed from the density distribution given in Fig. S14. 
The fact that the distribution of the phenol ring around the acid is very delocalized with no evident preferential arrangement close to the methyl group suggests weak interactions without directionality. This is observed both in classical simulations as well as in PIMD. Unlike the H-bond network, these apolar interactions are not associated with a clear structural CH-$\pi$ signature influencing the mixture.

\section{Conclusions}

This work demonstrates a framework that combines several traditional and data-driven atomistic modeling techniques to enable the simulation of complex, multi-component mixtures, taking as example the study of the solvation and acidity of \ce{CH3SO3H} and \ce{H2O2} in phenol.  

We use a neural-network machine-learning interatomic potential to reproduce accurate reference energetics based on dispersion-corrected hybrid DFT calculations. In order to obtain a robust model, we use semiempirical DFTB-D3H5 energies and forces as a baseline, achieving close-to-DFT accuracy (RMSE = 2 meV/Atom  and 146 meV/ \AA~ for the energies and forces respectively) while being 4000 times less expensive for the simulations we perform here. Using a multiple time step integrator with a direct MLP yields a further 3-fold speedup with no loss in accuracy. As a final component in our framework, we build a committee model to obtain accurate uncertainty estimation, that we use to monitor the error during simulations and for active learning. 
The robustness of our framework, combining MLP, accelerated sampling techniques, and on-the-fly uncertainty estimates, in turn, demonstrates a great potential to tackle reactions in condensed phase environments such as phenol hydroxylation.

We use this framework to study the solvation and the deprotonation of methanesulfonic acid in phenol, with and without the presence of a hydrogen peroxide molecule, using accelerated sampling techniques such as replica exchange molecular dynamics and metadynamics, and including quantum nuclear fluctuations using the PIGLET technique.
While the modeling of phenol hydroxylation catalyzed by CH$_{3}$SO${3}$H is in fact beyond the scope of this manuscript, the characterization of the acidity and solvation of a potential acid catalyst is a necessary prior step in the study of this reaction.

\ce{CH3SO3H} is found to be less acidic than protonated phenol and hydrogen peroxide, and to remain in its protonated state in unbiased simulations. The behavior is qualitatively different with respect to DFTB simulations, in which methanesulfonic acid readily loses its proton, which underscores the need for a MLP to reach hybrid DFT accuracy. 

An analysis of the solvation environment of \ce{CH3SO3H} reveals that the methanesulfonic hydroxyl group acts as a strong hydrogen bond donor toward phenol, while the sulfonyl oxygen atoms act as acceptors. Hydrogen peroxide binds strongly to the acid OH, competing with phenol and substituting it almost completely in simulations that include nuclear quantum fluctuations. Contrary to phenol, \ce{H2O2} interacts very weakly  with the sulfonyl oxygen atoms.
The highly structured environment, with directional hydrogen bonds and competition between phenol and hydrogen peroxide, underscore the need for an explicit, condensed-phase treatment of the system. These insights into the acid and \ce{H2O2} solvation shells also suggest that the reaction will never involve an isolated \ce{H3O2+} but rather an acid-\ce{H2O2} complex, which could potentially control the regioselectivity.
In addition to the hydrogen bond network, we also evaluated the importance of CH-$\pi$ interactions between the acid methyl group and the phenol ring but did not identify any relevant pattern influencing the mixture.

These simulations demonstrate that in order to obtain quantitative insights into the behavior of complex mixtures it is necessary to describe explicit solvation, achieve hybrid DFT accuracy using state-of-the-art MLPs, sample thoroughly thermal and quantum fluctuations of the nuclei, and analyze trajectories with automatic data-driven techniques. 
We have brought together all these ingredients into a robust, flexible framework that is readily applicable to the study of chemical reactions in complex condensed-phase environments.


\begin{acknowledgement}
The authors thank Venkat Kapil, Giulio Imbalzano, Federico Giberti, Piero Gasparotto and Raimon Fabregat for discussions. KR and VJ were supported by an industrial grant with Solvay. This project was initiated within the framework of the NCCR MARVEL, funded by the Swiss National Science Foundation (SNSF).

\end{acknowledgement}

\begin{suppinfo}
Supplementary Information (SI) available:   Neural network training and simulation set ups - Technical details - Error distribution of baselined training - MTS benchmarking -  Density distribution computations - Temperature dependence of oxygen-oxygen pair correlation functions from REMD simulation -  Hydrogen bonding network of the acid with hydrogen peroxide - NQEs in apolar interaction. See DOI: 10.1039/cXCP00000x/
\end{suppinfo}

\bibliography{bib}

\begin{tocentry}
\begin{center}
    \includegraphics[scale=0.225]{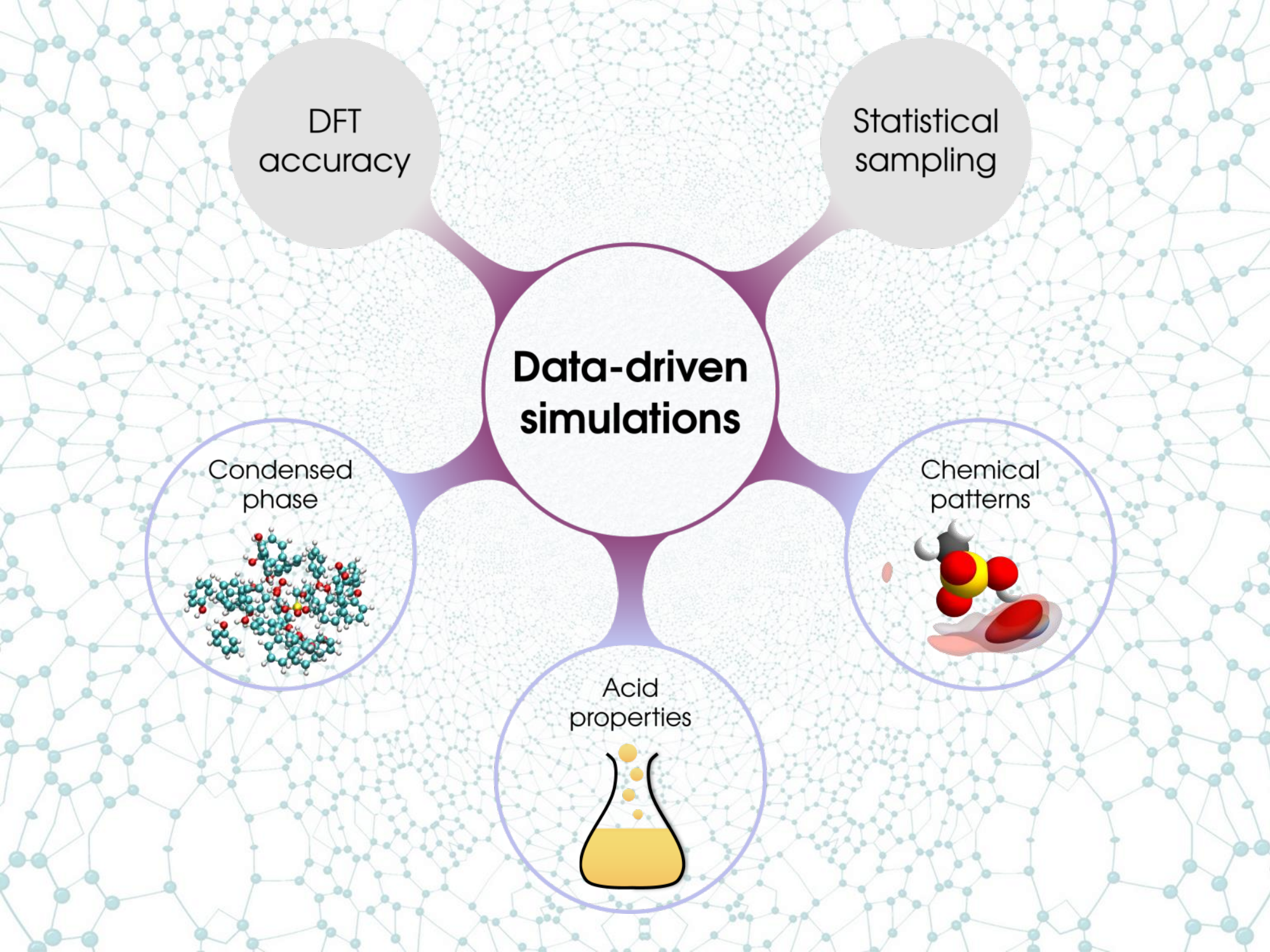}
\end{center}
\end{tocentry}

\end{document}